\documentclass[notoc,paper,letterpaper]{JHEP3} 
\usepackage{amssymb}

\newcommand{\be}{\begin{equation}}
\newcommand{\ee}{\end{equation}}
\newcommand{\bea}{\begin{eqnarray}}
\newcommand{\eea}{\end{eqnarray}}
\newcommand{\ie}{{\it i.e.}}
\newcommand{\eg}{{\it e.g.}}
\newcommand{\etc}{{\it etc.}}

\def\C{\mathbb{C}}
\def\I{\mathbb{I}}

\def\U{{\rm U}}
\def\GL{\mathop{\rm GL}}
\def\SU{\mathop{\rm SU}}
\def\SO{\mathop{\rm SO}}
\def\Sp{\mathop{\rm Sp}}
\def\tr{\mathop{\rm tr}}
\def\Pf{\mathop{\rm Pf}}
\def\t{\tilde}
\def\h{\hat}
\def\wt{\widetilde}

\def\bar{\overline}
\def\del{{\partial}}

\def\a{\alpha}

\def\b{\beta}
\def\tbe{\t\b}

\def\d{\delta}
\def\e{\epsilon}

\def\L{\Lambda}
\def\s{\sigma}

\def\nf{{N_f}}
\def\nc{{N_c}}

\def\tb{\wt b}

\def\tB{\t B}
\def\BB{{\cal B}}
\def\tBB{\t\BB }
\def\NN{{\cal N}}
\def\MM{{\cal M}}
\def\SS{{\cal S}}
\def\WW{{\cal W}}
\def\We{\WW_{\rm eff}}
\def\we{w_{\rm eff}}
\def\Mt{M^T}
\def\Wx{W_X}
\def\Wxt{\Wx^T}
\def\Xt{X^T}
\def\MMt{\MM^T}

\title{Generalized Konishi anomaly, Seiberg duality and 
singular effective superpotentials}

\author{Philip C. Argyres and Mohammad Edalati\\
Physics Department, University of Cincinnati, Cincinnati OH 45221-0011\\
\email{argyres,edalati@physics.uc.edu}}
\abstract{
Using the generalized Konishi anomaly (GKA) equations, we
derive the effective superpotential of four-dimensional 
$\NN=1$ supersymmetric $\SU(\nc)$ gauge theory with 
$\nf=\nc+2$ fundamental flavors.  We find, however, that
the GKA equations are only integrable in the Seiberg dual
description of the theory, but not in the direct
description of the theory.  The failure of 
integrability in the direct, strongly coupled, description 
suggests the existence of non-perturbative corrections
to the GKA equations.
}

\begin{document}

\section{Introduction}

Holomorphicity of the superpotential and gauge couplings, global 
symmetries and the weak-coupling limit enable one to obtain exact 
results in supersymmetric gauge theories (for reviews see 
\cite{is9509,p9702}), making these 
theories more tractable than their non-supersymmetric cousins. 
Since supersymmetric gauge theories exhibit a wealth of 
non-perturbative phenomena such as dynamically generated 
superpotentials with associated confinement or chiral symmetry 
breaking \cite{ads84,nsvz85}, deformed classical moduli spaces 
\cite{s9402}, Seiberg  duality \cite{s9411}, \etc., and since 
some of these phenomena also occur in non-supersymmetric theories, 
supersymmetric gauge theories are usually considered as a way to
qualitatively study non-perturbative aspects of ordinary gauge 
theories.  Therefore having a clear picture of the behavior of 
supersymmetric gauge theories may shed light on a better 
understanding of the dynamics of strongly-coupled gauge theories 
with no supersymmetry.

Despite much progress in the effective dynamics of four-dimensional 
$\NN=1$ supersymmetric QCD, the behavior of the effective 
superpotential for a number of flavors $\nf$ large compared to the 
number of colors $\nc$ is not well-understood.  This is because, 
firstly, when the number of flavors increases there are typically 
additional light degrees of freedom at the origin of the moduli 
space that one needs to include in the effective description. 
Secondly, the effective superpotentials become singular when 
expressed in terms of the local gauge-invariant light degrees 
of freedom away from the origin; more precisely, the potentials 
derived from such effective superpotentials have cusp-like 
singularities at their minima \cite{ae0510}.  Thirdly, the 
dependence of these effective 
superpotentials on the strong coupling scale of the theory $\L$ is 
such that they apparently diverge in the the weak coupling limit 
$\L\to 0$.  Because of these problems the physical meaning of such 
superpotentials is thought to be problematic. 

We have argued elsewhere \cite{ae0510} that effective superpotentials 
for the light gauge-invariant degrees of freedom away from the origin 
must nevertheless exist.  Furthermore, direct computation in the case 
of $\SU(2)$ superQCD shows \cite{ae0510} that these superpotentials,
although singular, are nevertheless physically sensible, and reproduce 
both the low energy physics as well as certain higher-derivative terms 
in an intrinsic description on the moduli space away from the origin 
\cite{bw0409}.  

In this paper we extend our arguments to $\SU(\nc)$ superQCD by 
computing its singular effective superpotential.  Unlike the $\SU(2)$ 
case, the $\SU(\nc)$ case has a smaller global symmetry group, making
it harder to find the superpotential.  We deal with this
by solving a system of differential equations for the
effective superpotential \cite{c0305} derived from the
generalized Konishi anomaly (GKA) equations \cite{cdsw0211}.
The complexity of this system increases with the number of
massless fundamental flavors $\nf$, but we
are able to solve them in the first interesting case,
$\nf=\nc+2$.  

However, there are some subtleties involved in applying
the GKA equations in this case:  the GKA equations are
not integrable when applied to the (strongly coupled) direct 
description of the theory, but are integrable when applied
to the (weakly coupled) Seiberg dual description of the theory.
In the rest of this section,
we will explain these subtleties in more detail and discuss the issues 
that they raise concerning the possible non-perturbative exactness
of the GKA equations.  We leave the technical details of the
calculations to subsequent sections.

The indirect argument for the existence of the effective
superpotential referred to above goes as follows:
Wilsonian effective superpotentials are assured to exist only 
if there is a region in the configuration space of the chosen 
chiral fields where all of them are light together and comprise 
all the light degrees of freedom. 
If this condition is satisfied, then the resulting effective 
superpotential can be extended over the whole configuration space 
by analytic continuation using the holomorphicity of the 
superpotential. 
For a large enough number of flavors the theory becomes IR free 
and the only region where all the components of the chosen chiral 
vevs become light at the same time is at the origin.  
We know what the light degrees of freedom are near that point since 
we have a weakly coupled lagrangian description there. 
The physics can be made arbitrarily weakly coupled simply by taking 
all scalar field vevs $\langle\phi\rangle << \L$ where $\L$ 
is the strong coupling scale (or UV cutoff) of the IR free theory.  
In this limit the physics is just the classical Higgs mechanism, and 
all particles get masses of order $\langle\phi\rangle$ or less.  
The Wilsonian effective description results from integrating out modes 
with energies greater than a cutoff, which we take to be some multiple 
of $\langle\phi\rangle$.  
The effective action will then include all local gauge-invariant 
operators made from the fundamental fields in the lagrangian and which 
can create particle states with masses below the cutoff.  
For the purpose of constructing the effective superpotential, the 
relevant local gauge-invariant operators are those in the chiral ring. 
It is then just a matter of constructing in the classical gauge theory
a set of operators which generate the chiral ring.  We will refer to 
this set as the classical chiral operators of the theory.

In a weakly coupled $\SU(\nc)$ superQCD a basis of local gauge-invariant 
operators in the chiral ring (the classical chiral operators) is comprised 
of just the glueball, meson, and baryon operators \cite{cdsw0211,s0212}.  
An effective superpotential which is a function of these operators 
must then exist.  For $\nf>\nc+1$, the quantum moduli space is also 
the same as the classical one \cite{s9402}, but effective 
superpotentials (singular or not) for these cases have not been
found before.  Also, this theory has an equivalent description in
the IR in terms of a ``Seiberg dual" $\SU(\nf-\nc)$ supersymmetric 
QCD with $\nf$ (dual) fundamental quarks and anti-quarks and a set 
of singlet scalars coupled to the dual mesons through a superpotential 
\cite{s9411}. 

Note that the above argument does not directly show the existence of 
such effective superpotentials in the asymptotically free case.  
In particular, for theories in the ``conformal window" where neither 
the direct nor Seiberg dual description is IR free (${3\over2}\nc <
\nf<3\nc$ for $\SU(\nc)$ gauge group), we have no useful description 
of the light degrees of freedom at the origin of moduli space. 
Nevertheless, given an effective superpotential for an IR free theory, 
one can then successively add mass terms to the effective superpotential 
and integrate out massive flavors to derive consistent effective 
superpotentials in the conformal window. 
This then assures us that effective superpotentials exist for all 
numbers of light flavors in supersymmetric QCD. 

Our method for deriving the effective superpotential for the
$\nf>\nc+1$ theory will be to integrate the generalized Konishi 
anomaly (GKA) equations \cite{cdsw0211} following the approach
of \cite{binor0303,c0305}.  
The resulting equations become very complicated \cite{c0305} for 
large numbers of flavors, so we are not able to solve them directly
in the IR free case $\nf\geq3\nc$, and then integrate out flavors as 
in the above argument.

For $\nf=\nc+2$, however, the GKA equations simplify to a first order 
matrix differential equation simple enough that we can analyze it.  
We show that the GKA equations for the effective superpotential 
are not integrable in this case for $\nc>2$.  
This is not in direct conflict with the general arguments advanced 
above:  for $\nf=\nc+2$ and $\nc>2$, the theory at the origin is 
strongly coupled in terms of its microscopic fields, so an effective 
description in terms of the chiral ring operators made from these 
fields simply need not exist.  

However, this failure of integrability presents a sharper puzzle 
in light of the following:  we can nevertheless calculate an effective 
superpotential by using the fact that for $\nf=\nc+2$ with $\nc\ge4$ 
the Seiberg dual description is an IR free $\SU(2)$ gauge theory.  
By applying the GKA equations to the Seiberg dual description we 
derive the effective superpotential of the theory in terms of the 
dual chiral fields.  It is given in equation (\ref{weffws}) below,
where we have used the map \cite{s9411} between direct and dual 
chiral operators to interpret this as an effective 
superpotential in terms of the classical chiral operators of 
the direct theory---the mesons $M^i_j$, baryons $B_{ij}$ and 
$\tB^{ij}$, and the glueball $S$.

We have thus found the effective superpotential in terms of the
classical chiral operators.  This raises the question of why 
were the GKA equations not integrable in the direct theory in the
first place?  We interpret this failure of integrability as 
indicating that the GKA equations get non-perturbative corrections.  
It remains to characterize more precisely the nature of the 
non-perturbative corrections to the GKA equations.
One possibility is that there exists a non-perturbatively 
modified set of GKA equations in terms of the classical chiral 
operators (\ie, glueball, mesons, and baryons in our case).
Another possibility is that there are additional 
operators in the chiral ring which are independent of the 
classical chiral operators; when included, they could render
the GKA equations integrable, in much the same way that the
extra singlet field in the Seiberg dual description does.
It has not been ruled out that such additional fields could 
be seen in the semi-classical description as higher derivative 
chiral operators.\footnote{We thank S. Hellerman for discussions
on this point.}  (Note that the non-trivial higher-derivative 
chiral operators constructed in \cite{bw0409} are {\em not} 
candidates, since they are $\bar Q$-exact when extended off the 
moduli space to the configuration space of chiral vevs.)  Finally, 
both possibilities---a non-perturbative deformation of the
GKA equations and the inclusion of additional chiral 
fields---could occur together.

Once the issue of non-perturbative corrections to the GKA equations is 
raised, it applies equally well to the GKA equations derived in the 
Seiberg dual description.  It is an open question whether the effective 
superpotential derived below in the dual description is correct or not 
for $\nf\ge6$.  For even though, when $\nf\ge6$, the dual description is 
weakly coupled at the origin of moduli space, it becomes strongly coupled 
an arbitrarily small distance away from the origin since the superpotential 
term in the dual theory destabilizes the free fixed point at the origin 
\cite{s9411}.

The remainder of the paper carries out the computations described
qualitatively above, and is organized as follows.  To illustrate the 
method in a simple case first, and for later comparison to the Seiberg 
dual description of the $\SU(\nc)$ case, in section 2 we consider the 
case of $\SU(2)$ gauge group with $\nf\geq 4$ flavors.   We show how 
the GKA equations written in terms of $\SU(2)$ mesons and baryons gives 
an effective superpotential matching that found in \cite{ae0510} where 
we worked instead with the single antisymmetric meson field appropriate 
to an $\Sp(1)$ description of the theory (\ie, one which makes the 
enlarged global symmetry group of the $\SU(2)$ compared to the general 
$\SU(\nc)$ theory manifest).  In section 3 we apply the GKA equations 
to $\SU(\nc)$ superQCD with $\nf=\nc+2$ and $\nc\ge4$.  We show that the
resulting equations for the effective superpotential are not integrable 
for $\nf\ge6$, but that they are integrable and match the $\SU(2)$ result 
of the previous section for $\nf=4$.  

In section 4 we apply the GKA equations to the Seiberg dual of the theory 
in section 3, and solve for the effective superpotential.   The form of 
this superpotential is complicated: integrating out the heavy glueball 
gives an effective superpotential of the form $\sqrt{\det M} f(X)$ where 
$X=M\tB\Mt B/\det M$, but a closed-form expression for $f(X)$ is not found. 
Instead, we show that $f$ obeys a nonlinear first order matrix differential 
equation (\ref{mde}).  A power series 
expansion of the solution to order $X^4$ is computed in (\ref{weffexp}).
We then compare this result to the $\SU(2)$ effective superpotential of 
section 2 when $\nf=4$, and show that they agree at least to order 
$X^4$.

\section{$\We$ for $\bf\nf\ge4$ SU(2) superQCD}

We show how to calculate the effective superpotential of 
$\SU(2)$ superQCD with $\nf\geq4$ using the generalized
Konishi anomaly equations, following \cite{binor0303,c0305}.

$\SU(2)$ superQCD has $\nf$ massless quark and anti-quark 
chiral multiplets, $Q^i_a$ and $\t Q^a_i$, transforming 
in the $\bf 2$ and $\bf{\bar 2}$ of the gauge group, respectively.  
Here $a=1,2$ is the color index and $i=1,\ldots,\nf$ the flavor 
index.  
The apparent global symmetry of the theory is $\SU(\nf) \times 
\SU(\nf) \times \U(1)_B \times \U(1)_R$. 
Since $\bf 2$ and $\bf{\bar 2}$ are equivalent representations, 
though, $\SU(2)$ superQCD actually has the larger symmetry group
$\SU(2\nf)\times \U(1)_R$, which is large
enough to determine the effective superpotential uniquely
\cite{ae0510}.  Here, since we are looking to the generalization
to $\SU(\nc)$ which only has the smaller symmetry group, we will
analyze the $\SU(2)$ case in terms of the $Q$'s and $\t Q$'s 
keeping only the $\SU(\nf)\times\SU(\nf)\times\U(1)_B\times\U(1)_R$ 
symmetry manifest.

The classical moduli space is parameterized by the vevs of the meson 
$M$ and the baryon $\BB$, $\t\BB$ chiral superfields defined by
\be\label{2.2}
M^i_{\ j} := Q^i_a \t Q^a_j ,\qquad
\BB^{ij} := \e^{ab} Q^i_a  Q^j_b ,\qquad
\t\BB_{ij} := \e_{ab} \t Q^a_i \t Q_j^b,\qquad
S :=\tr(W^\a W_\a)/(32\pi^2),
\ee
where we have also defined the glueball chiral superfield $S$.  
These fields can be assigned the charges $R(S)=2$ and $R(M)=R(B)
=R(\tB) =2(\nf-2)/\nf$ under the non-anomalous $\U(1)_R$ symmetry. 
The meson and baryon vevs cannot take arbitrary values but are 
subject to constraints following from (\ref{2.2}),
\be\label{2.6}
\BB^{[ij}M^{k]}_{\ \ell} = M^i_{\ [j} \t\BB_{k\ell]} 
= \BB^{[ij}\BB^{k]\ell} = \t\BB_{i[j} \t\BB_{k\ell]} 
= M^{[i}_{\ k} M^{j]}_{\ \ell} - \BB^{ij}\t\BB_{k\ell} = 0,
\ee
where the square brackets denote antisymmetrization.  These 
constraints imply that $M$, $\BB$, and $\t\BB$ all have rank 
less than or equal to 2 and, up to flavor rotations, take the form
\be
M = \pmatrix{m_1&&\cr &m_2&\cr &&{\bf 0}},\qquad
\BB = \pmatrix{&b&\cr -b&&\cr &&{\bf 0}},\qquad
\t\BB = \pmatrix{&\tb&\cr -\tb&&\cr &&{\bf 0}},
\ee
with $m_1m_2=b\tb$ and $\bf 0$ the $(\nf-2)\times(\nf-2)$ matrix of
zeros.  For $\nf \ge 3$, the quantum moduli space is also the same as 
the classical one \cite{s9402}. 

For $\SU(2)$ superQCD with fundamental flavors $M$, $\BB$, $\t\BB$, 
and $S$ are thought to generate all non-trivial local gauge-invariant 
operators in the chiral ring of the classical theory \cite{cdsw0211,s0212}. 
When $\nf=4$ or $5$ the theory is strongly coupled, and has new 
massless degrees of freedom at the origin of moduli space, so 
the chiral ring might be deformed or enlarged from the classical
answer.  But for $\nf\ge6$, where the theory is IR free, the classical
description is as accurate as we like (in the vicinity of the origin
of the moduli space).  So we will make the assumption that we can 
write our effective superpotential in terms of just $S$, $M$, $\BB$, 
and $\t\BB$.

However, for $\nf\ge4$, the global symmetries allow infinitely
many terms in the effective superpotential, making it hard
to guess its correct form.  So, instead, we use the generalized 
Konishi anomaly (GKA) equations to derive the effective superpotential. 
If $F^i_r(\Phi, W_\a)$ are holomorphic functions transforming 
in the same representation of the gauge group as a chiral
superfield $\Phi^i_r$ ($i$ is a flavor index and $r$ an index for
the gauge representation), then the GKA equation \cite{cdsw0211} is
\be\label{2.8}
\left\langle {\del\WW_{\rm tree}\over\del\Phi^j_r} F^i_r\right\rangle =
{1\over32\pi^2} \left\langle (W^\a W_\a)^s_t 
{\del F^i_s\over\del\Phi^j_t}\right\rangle ,
\ee
which can be interpreted as the anomalous Ward identity 
coming from the field transformation $\d\Phi^i_r= F^i_r$. 
Here $\WW_{\rm tree}$ is the classical superpotential. 
The GKA equation is perturbatively 
one-loop exact \cite{cdsw0211}.  It has also been shown 
\cite{s0311} that it does not get non-perturbative corrections 
for a $\U(N)$ gauge theory with matter in the adjoint representation 
as well as for $\Sp(N)$ and $\SO(N)$ gauge theories with matter 
in symmetric or antisymmetric representations.  
For the theories we are discussing here, its non-perturbative
status is not known; however, as we show below, there is strong
evidence, at least for $\SU(2)$, that the GKA equations are
actually non-perturbatively exact.

Consider now $\SU(2)$ superQCD with the classical superpotential
\be\label{2.11}
\WW_{\rm tree}=m^i_{\ j}(\h M^j_{\ i}-M^j_{\ i})+ 
b_{ij}({\h \BB}^{ij}-\BB ^{ij})+\tb^{ij}(\h {\tBB}_{ij}
-{\tBB}_{ij}).
\ee
Here $m^i_{\ j}, b_{ij}$ and $\tb^{ij}$ are Lagrange multipliers 
constraining the operators ${\h M}^j_{\ i}$, ${\h\BB}^{ij}$ 
and ${\h{\t\BB}}_{ij}$ to have $M^j_{\ i}$, $\BB^{ij}$ 
and ${\t\BB}_{ij}$ as their vacuum expectation values, 
respectively.   (Whenever we need to distinguish an operator
from its vev, we put a hat on the operator.)  We are looking
for the effective superpotential $\We$ as a function of
the vevs $S$, $M$, $\BB$, and $\t\BB$.  It follows from
(\ref{2.11}) and the nature of the Legendre transform
\cite{ils9403,i9407,is9509} that
\be\label{2.12}
m^i_{\ j} = -{\del\We\over\del M^j_{\ i}},\qquad
b_{ij} = -{1\over2}{\del\We\over\del\BB^{ij}},\qquad
\tb^{ij} = -{1\over2}{\del\We\over\del{\t\BB}_{ij}},
\ee
where the factors of 2 come from the antisymmetry of the baryons.

We now use the GKA equations to 
determine the dependence of the Lagrange multipliers on
$M$, $\BB$, $\t\BB$, and $S$.  First set $F^i_r = \Phi^i_r = 
Q^i_a$ in (\ref{2.8}) yielding
\be\label{2.13}
M m = S +2\BB b, 
\ee
where we are using a matrix notation on the flavor indices
(so that, \eg, the last equation stands for $M^i_{\ k} m^k_{\ j} 
= S{\d}^i_{\ j} +2\BB^{ik}b_{kj}$).  A similar equation,
\be\label{2.14}
 mM=S+2{\tb}{\tBB}, 
\ee
follows from taking $F^i_r = \Phi^i_r = \t Q^a_i$.
Two more independent equations follow from taking
$F^i_r=\e_{ab}{\t Q}^b_i$ and $\Phi^i_r=Q^i_a$, and
from taking $F^i_r=\e^{ab} Q^i_b$ and $\Phi^i_r=\t Q^a_i$,
giving
\be\label{2.15}
\t\BB m =  -2 \Mt b ,\qquad
m \BB =  -2 \tb \Mt. 
\ee

We will carry out subsequent calculations at a generic point 
on the configuration space of $S$, $M$, $\BB$ and $\t\BB$ where they 
are all invertible matrices.  Note, however, that when $\nf$ is 
odd, $\BB$ and $\t\BB$, being odd rank antisymmetric matrices,
are never invertible.  We get around this problem by restricting
ourselves to an even number of flavors only.  Once we have found the 
superpotential for even $\nf$'s, we can add a mass term for one 
flavor and integrate it out to get the effective superpotential 
for the odd $\nf-1$ flavors. 

Therefore, multiplying (\ref{2.13}--\ref{2.15}) by appropriate inverses
and substituting for the Lagrange multipliers $m$, $b$, and $\tb$ using
(\ref{2.12}) gives a set of partial differential equations for $\We$
\bea\label{2.21}
{\del\We\over\del\BB^{ij}} &=&
S\left[ (\t\BB M^{-1}\BB+\Mt)^{-1}\t\BB M^{-1}\right]_{ij},\nonumber\\
{\del\We\over\del{\t\BB}_{ij}} &=&
S\left[ M^{-1}\BB(\t\BB M^{-1}\BB+\Mt)^{-1}\right]^{ij},\nonumber\\
{\del\We\over\del M^j_{\ i}} &=&
S\left[ M^{-1}\BB(\t\BB M^{-1}\BB+\Mt)^{-1}\t\BB M^{-1}
-M^{-1}\right]^i_{\ j}.
\eea
Integrate the first equation in (\ref{2.21}) to find
\be\label{2.22a}
\We = -{S\over2}\ln\det\left(\I + M^{-T}\t\BB M^{-1}\BB\right) 
+G(M,\t\BB,S),
\ee
where $M^{-T}=(\Mt)^{-1}$, $\I$ is the $\nf\times\nf$ identity
matrix, and $G$ is an undetermined integration function.  Comparing 
the second equation in (\ref{2.21}) with the derivative of (\ref{2.22a}) 
with respect to $\t\BB$ gives $\del G/\del\t\BB = 0$.  Also, comparing
the derivative of (\ref{2.22a}) with respect to $M$ to the third 
equation in (\ref{2.21}) gives $\del G/\del M=-S M^{-1}$, so that
\be\label{2.22e}
G = -S \ln\det(M/\L^2) + H(S),
\ee
for some undetermined function $H(S)$.  The $\L$-dependence was 
determined by dimensional analysis, where $\L$ is the strong-coupling
scale of the $\SU(2)$ superQCD.  

Equivalently, the global flavor symmetry implies that $\We=\We(X, 
\det M, S)$ where $X:=M^{-T}\t\BB M^{-1}\BB$.  Plugging this functional 
form into (\ref{2.12}--\ref{2.15}) gives simple matrix differential 
equations leading to (\ref{2.22a}--\ref{2.22e}).

$H(S)$ is determined up to a constant by the $\U(1)_R$
symmetry.  Since $R(\We)=2$, $H$ must be linear in $S$, plus a 
logarithmic piece to cancel the $\U(1)_R$ transformation of the
$-S\ln\det M$ term, giving
\be
H(S) = (2-\nf) S [\a-\ln(S/\L^3)]
\ee
for some undetermined constant $\a$.  We can determine $\a$ by 
matching to the Veneziano-Yankielowicz superpotential \cite{vy82}, 
$W_{\rm VY}(S)=2S[1-\ln (S /\L_{\rm YM}^3)]$, for pure $\SU(2)$ 
superYangMills.  It is a short exercise to integrate out the mesons 
and baryons in (\ref{2.22a}) and match strong coupling scales to 
find $\a=1$.  We therefore find that the effective superpotential is
\be\label{2.22}
\We = -{S\over2} \ln\left[(\det M)^2 \det(\I+M^{-T}\t\BB M^{-1}\BB)\right] 
+ (2-\nf) S (1-\ln S)
+ (6-\nf) S\ln\L.
\ee
Since $S$ is massive we can integrate it out by solving its equation 
of motion, $\del\We/\del S=0$ to find 
\be\label{2.23}
\We(M,\BB,\t\BB) = (2-\nf) \left[\L^{\nf-6} \det M
\sqrt{\det(\I+M^{-T}\t\BB M^{-1}\BB)} \right]^{1/(\nf-2)}. 
\ee 

This superpotential reproduces all known low energy aspects of
$\SU(2)$ superQCD.  The easiest way to see this is to convert
it to a description which makes the full global flavor symmetry
manifest.  As we mentioned earlier, $\SU(2)$ superQCD with $\nf$ 
fundamental $Q^i_a$ and $\nf$ anti-fundamental ${\t Q}^a_i$ 
can be equivalently described in terms of $2\nf$ doublets
${\cal Q}^I_a$, $I=1,\cdots, 2\nf$ with ${\cal Q}^i_a = Q^i_a$ and
${\cal Q}^{\nf+i}_a = \e_{ab}{\t Q}^b_i$.  Hence $M$, $\BB$ and $\t\BB$ 
are combined into an antisymmetric $2\nf \times 2\nf$ matrix $V^{IJ}
= \e^{ab}{\cal Q}^I_{a} {\cal Q}^J_{b}$,
\be\label{b4.2}
V=\pmatrix{\BB&M\cr -M^T&\t\BB\cr}.
\ee
After a bit of algebra\footnote{Write $V={\BB\ 0\choose0\ \tBB}
{1\ x\choose y\ 1}$ with $x:=\BB^{-1}M$, $y:=-\tBB^{-1}\Mt$.  Use the 
identity $\det{1\ x\choose y\ 1}=\det(1-xy)$, so $\det V = \det\BB 
\det\tBB \det(1-xy) = (\det M)^2 \det(-y^{-1} x^{-1}+1)$, which gives
(\ref{2.23}).} it is seen that our singular effective 
superpotential (\ref{2.23}) can be written in terms of this 
new variable as
\be\label{b4.4}
\We = (2-\nf)\left(\L^{\nf-6}\sqrt{\det V}\right)^{1/(\nf-2)},
\ee
making the $\SU(2\nf)$ global symmetry manifest.  Indeed, (\ref{b4.4})
coincides with the singular effective superpotential found in
\cite{ae0510}, and so it satisfies all the checks discussed there:
it gives rise to the correct moduli space, is consistent under
integrating out flavors, and reproduces all the higher-derivative
F-terms found in \cite{bw0409}.

The success of this calculation can be taken as evidence that the 
GKA equations are non-perturbatively exact
for $\SU(2)$ superQCD.

\section{$\We$ for $\bf\nf=\nc+2$ superQCD: 
non-integrability of GKA equations}

$\SU(\nc)$ superQCD has $\nf$ massless quark chiral fields 
$Q^i_a$ and $\nf$ massless anti-quark chiral fields ${\t Q}^a_i$ 
transforming in the fundamental and anti-fundamental representations, 
respectively.  Here $i=1,\ldots, \nf$ is the flavor index and 
$a=1,\ldots, \nc$ is the color index.   When $\nf=\nc+2$ the classical 
moduli space is parameterized by the gauge-invariant vevs of the glueball,
meson, and baryons defined by
\bea\label{4.1}
\h S & := & {1\over32\pi^2} \tr(W^\a W_\a) ,\nonumber\\
\h M^i_j & := & Q^i_a \t Q_j^a ,\nonumber\\
\h B_{ij} & := &  {1\over\nc!} \e_{ijk_1\cdots k_\nc}
\e^{a_1 \cdots a_\nc} Q^{k_1}_{a_1}\cdots Q^{k_\nc}_{a_\nc}, \nonumber\\
\h {\tB}^{ij} & := & {1\over\nc!} \e^{ijk_1\cdots k_\nc}
\e_{a_1 \cdots a_\nc}\t Q_{k_1}^{a_1}\cdots\t Q_{k_\nc}^{a_\nc}.
\eea
The global symmetry of the theory is $\SU(\nf)\times\SU(\nf)
\times\U(1)_B\times\U(1)_R$.  The $\U(1)_R$ charges are $R(S)=2$, 
$R(M)=4/\nf$, and $R(B)=R(\tB)=2\nc/\nf$.  The 
classical moduli space is described by the constraints that $M$, 
$B$, and $\tB$ satisfy by virtue of their definitions,
\be\label{4.4}
B_{ik} M^k_j = M^i_k \tB^{kj}= B_{[ij}B_{k]\ell}
= \tB^{[ij}\tB^{k]\ell} = \tB^{ij} B_{k\ell} - 
{M^{-1}}^{[i}_k {M^{-1}}^{j]}_\ell\ \det M = 0.
\ee
Square brackets denote antisymmetrization; antisymmetrization on $n$ 
indices consists of $n!$ terms (\ie, with out a factor of $1/n!$).

They imply that by appropriate flavor rotations $M$, $B$, and 
$\tB$ can be put in the form
\be\label{cc2}
M = \pmatrix{\bf m&&\cr &&\phantom{b}\cr &\phantom{-b}&},\qquad
B = \pmatrix{\phantom{\bf m}&&\cr &&b\cr &-b&},\qquad
\tB=\pmatrix{\phantom{\bf m}&&\cr &&\tb\cr &-\tb&},
\ee
where $\bf m$ is an $\nc\times\nc$ matrix and $b$, $\tb$ are 
numbers satisfying $b\tb = \det({\bf m})$.  The classical and 
the quantum moduli spaces are the same \cite{s9402}, but at the 
origin there are extra light degrees of freedom.  At points away 
from the origin, the only light degrees of freedom are components 
of $M$, $B$, and $\tB$.  At the origin, $\SU(\nc)$ supersymmetric 
QCD with $\nf=\nc+2$ is an interacting superconformal field theory 
for $\nc=2$ and $3$.  For $\nc\geq4$ it becomes strongly coupled, but 
has an IR free dual description in the IR \cite{s9411}.

In this section we will try to construct an effective superpotential 
in terms of these fields which correctly describes the moduli space 
of vacua for points away from the origin.  As in the last section, we 
will use the GKA equations to systematically derive $\We$.  In fact, 
the GKA equations were used in \cite{c0305} to construct a set of coupled 
partial differential equations for the effective superpotentials of 
$\SU(\nc)$ supersymmetric QCD.  They have been integrated \cite{c0305} 
for $\nf=\nc$ and $\nf=\nc+1$ where the results are in agreement with 
those in \cite{s9402}.  Unfortunately, the GKA equations are quite 
complicated for $\nf\geq\nc+2$ flavors.  We will show below how to
simplify the GKA equations when $\nf=\nc+2$.

We briefly recap the derivation of the equations for $\We$ from the
GKA equations \cite{c0305}.  The strategy is the same as in the 
$\SU(2)$ case discussed in the last section: start with the tree 
level superpotential
\be\label{4.5}
\WW_{\rm tree}=m^i_j(\h M^j_i-M^j_i)+ 
b^{ij}(\h B_{ij}-B_{ij})+\tb_{ij}(\h{\tB^{ij}}-\tB^{ij})
\ee
where 
\be\label{4.6}
m^i_j=-{\del\We\over\del M^j_{i}},\qquad
b^{ij}=-{1\over 2}{\del\We\over\del B_{ij}},\qquad
\tb_{ij}=-{1\over 2}{\del\We\over\del\tB^{ij}},
\ee
are Lagrange multipliers enforcing that $M^j_i$, $B_{ij}$ and 
$\tB_{ij}$ be the vevs of the meson, baryon, and anti-baryon 
operators, respectively.  There is no need to introduce a Lagrange 
multiplier for $S$ because we are considering points away from the 
origin and $S$ is massive for these points.
We get two relations among the Lagrange multipliers by taking
$F_r$ to be the quark or antiquark field in the GKA 
equation (\ref{2.8}), and two more by taking it to be proportional
to $\e^{ijk\ell_2\cdots\ell_\nc} \e_{a a_2\cdots a_\nc} 
\t Q_{\ell_2}^{a_2}\cdots\t Q_{\ell_\nc}^{a_\nc}$, and similarly
with the $Q$'s.  The resulting GKA equations are
\bea\label{4.7}
M^i_k m^k_j &=& (S+b^{k\ell}B_{\ell k})\d^i_j 
- 2b^{ik}B_{kj} , \nonumber\\
m^i_k M^k_j &=& (S+\tb_{k\ell}\tB^{\ell k})\d^i_j 
- 2\tb^{ik}\tB_{kj} , \nonumber\\
m^{[j}_i\tB^{k\ell]} &=& 2b^{hg} 
{M^{-1}}^{[j}_g {M^{-1}}^k_h {M^{-1}}^{\ell]}_i \det M , \nonumber\\ 
m^i_{[j} B_{k\ell]} &=& 2\tb_{hg} 
{M^{-1}}_{[j}^g {M^{-1}}_k^h {M^{-1}}_{\ell]}^i \det M .
\eea
Note that the right sides of the last two equations, though they
are written using $M^{-1}$, are actually polynomial in $M$.

Now, the global flavor symmetry implies\footnote{This symmetry argument 
is not entirely straightforward.  The $\U(1)_B$ baryon number symmetry 
implies that for each $B$ there must be an accompanying $\tB$ in 
each term.  Since $\We$ is an $\SU(\nf)\times\SU(\nf)$ singlet, all 
the flavor indices must be contracted in each term.  Contractions 
with the totally antisymmetric epsilon tensors can always be reduced 
to products of $\det M$ and $\Pf B\cdot\Pf\tB$.  The only other way 
to contract indices of $B$ and $\tB$ is with an $M$ as $BM\tB$ 
(or its transpose), and since these in turn must be contracted, 
another factor of $M$ must be included.  There are four ways of 
doing this---$\Mt BM\tB$ and its three cyclic permutations---but 
upon making a flavor singlet expression a trace must be taken, so 
the cyclic order does not matter.  Finally, the product of Pfaffians 
of baryons is not independent of $X$ and $\det M$, since $\Pf B\cdot
\Pf\tB = \sqrt{\det(\Mt BM\tB)}/\det M$.

Alternatively, one can derive this directly from the GKA equations.
Use them to deduce (\ref{4.10}) and a similar relation for $b$, 
multiply these by $\tB$ and $B$, respectively, then substitute the second
into the first.  One finds that $\tb\tB$ depends on $B$ and $\tB$
only through $X$.  Since $\tb\sim\del\We/\del\tB$, it follows that 
the dependence of $\We$ on $B$ and $\tB$ is solely through $X$.

Note that this symmetry argument is no longer effective when $\nf>\nc+2$.
For then $B$ and $\tB$ have more than two indices, and the analog
of $X$ is no longer a matrix, but has $\nf-\nc-1$ upper and $\nf-\nc-1$
lower (antisymmetrized) indices.  These objects can be contracted
in many inequivalent ways to make flavor singlets.  This is the source
of the difficulty in integrating the GKA equations for $\nf>\nc+2$.}
that 
\be\label{X}
\We=S\,f(X,S^{-2}\det M),
\qquad\mbox{where\ \ }
X := {\Mt BM\tB\over\det M} ,
\ee
and we are using matrix notation for the meson and baryon 
fields.  The first two equations in (\ref{4.7}) then imply
\be\label{4.11}
\We= S\,W(X) - S\ln\det M + 2S(\ln S-1),
\ee
where $W$ is to be determined.  

The GKA equations (\ref{4.7}) imply a matrix differential equation 
for $W(X)$ as follows.  Contract the $ijk\ell$ indices in the
last equation in (\ref{4.7}) with three $M$'s, giving
$2(2-\nf)\tb\,\det M=\tr(mM) \Mt BM + \Mt m^T\Mt BM-\Mt BMmM$.
Substitute for $Mm$ using the first equation to get
\be\label{4.10}
2\,\tb\,\det M = -  \left[S+{\nc-2\over\nc}\tr(bB)\right] \Mt BM 
- {4\over\nc}\Mt BbBM .
\ee
Derivatives of (\ref{4.11}) with 
respect to $B$ and $\tB$ together with (\ref{4.6}) imply
\bea\label{4.12}
2\,b \,\det M &=& S\, (M\tB\Wx\Mt + M\Wxt\tB\Mt), \nonumber\\ 
2\,\tb\,\det M &=& S\, (\Wx\Mt BM + \Mt BM\Wxt),
\eea
where $\Wx:=\del W/\del X$.  Work at a generic point in parameter 
space where $M$, $B$, $\tB$ and $X$ are all invertible matrices.  
This is only possible if $\nf$, and therefore $\nc$, is taken to be 
even, as in the discussion in the previous section.  

Substitute (\ref{4.12}) in (\ref{4.10}) and multiply on both left 
and right by $\tB$ to obtain
\be\label{4.16}
\tB G(X) =-{G^T}(X) \tB, 
\ee
where 
\be\label{4.17}
G(X):=\Wx X + {1\over2}\left[1+{\nc-2\over\nc}
\tr(\Wx X)\right] X +{2\over\nc} X \Wx X,
\ee
and we have used that $\tB X$ is antisymmetric.  On the other hand, 
from the definition of $X$ (\ref{X}) it follows that $\tB X =\Xt\tB$, 
which implies
\be\label{4.17a}
\tB\,G(X)=G^T(X)\,\tB
\ee
since $G$ is a function of $X$ alone.  (\ref{4.17a}) and (\ref{4.16}) 
imply $G(X)=0$, which, after being multiplied from the right by $X^{-1}$, 
reads   
\be\label{4.18}
\Wx + {1\over2}\left(1+{\nc-2\over \nc}
\tr(\Wx X)\right)\I+{2\over\nc} X \Wx =0,
\ee
where $\I$ is the $\nf \times \nf$ identity matrix.  This is
the matrix differential equation for $W(X)$.

The trace of (\ref{4.18}), $\tr(\Wx X)=-[2\tr(\Wx)+\nc+2]/\nc$, 
allows us to eliminate $\tr(\Wx X)$ from (\ref{4.18}), giving
$\nc\Wx = (\nc+2X)^{-1}[(\nc-2)\tr(\Wx)-2]$.  The 
trace of this equation allows us to eliminate $\tr(\Wx)$ in turn,
giving the following differential equation for $W(X)$:
\be\label{4.21}
\Wx = {2(\nc +2X)^{-1} \over (\nc-2)\tr((\nc + 2X)^{-1})-\nc}.
\ee
When $\nc\neq 2$, we define the matrix $Y:=(\nc-2)(\nc+2X)^{-1}$, 
and substitute it into (\ref{4.21}) to obtain  
\be\label{4.22}
{\del W\over\del Y^j_k} Y^i_k = {1\over\nc-\tr(Y)} \d^i_j,
\ee
where we have explicitly written the indices to avoid any confusion. 
This differential equation is not integrable, as it is easy to check
that $\del^2 W/\del Y^k_l\del Y^i_j \neq \del^2 W/\del Y^i_j\del Y^k_l$.
This shows that the GKA equations for $\We$
are not integrable for even values of $\nc>2$. 

\subsection{Comparison to the SU(2) solution.}

For $\nc=2$ we are integrating the same GKA equations as we did
in section 2, though in terms of the baryons and anti-baryons $B$
and $\tB$ instead of their Hodge duals $\BB$ and $\t\BB$.  Thus
the GKA equations must be integrable in this case, and, of course,
give the same answer we found in section 2, namely, equation 
(\ref{2.23}) with $\nf=4$.  However, there is a subtlety in comparing 
these two computations, which we will now explain.  It will play an 
important part in our discussion of the results of integrating the 
Seiberg dual GKA equations in the next section.

For $\nc=2$, (\ref{4.18}) is indeed integrable, and integrates to 
give $W(X) = -{1\over2}\tr\ln(1+X)$.  Integrating out $S$ from
(\ref{4.11}) then gives 
\be\label{su2weff1}
\L{\cal W}_{\rm eff, SU(2)} = -2\sqrt{\det M} 
\det (1+X)^{1/4},
\ee
where $\Lambda$ is the strong coupling scale of the gauge theory.
On the other hand, the $\SU(2)$ effective superpotential for $\nf=4$
found in section 2 is
\be\label{su2weff2}
\L{\cal W}_{\rm eff, SU(2)} = -2 \sqrt{\det M}
\det (1+M^{-T}\t\BB M^{-1} \BB )^{1/4}.
\ee
Since the $\SU(2)$ baryon fields $\BB$ and $\t\BB$ as
defined in (\ref{2.2}) are Hodge-dual to the $\SU(\nc)$ baryons
$B$, $\tB$ defined in (\ref{4.1}), and since rank$(M)=\nf=4$,
it follows that
\be
M^{-T}\t\BB M^{-1} \BB = {1\over2} \tr(X)- X ,
\ee
so that the effective superpotential reads
\be\label{su2weff3}
\L{\cal W}_{\rm eff, SU(2)} = -2 \sqrt{\det M}
\det (1+ \textstyle{1\over2} \tr(X)- X)^{1/4}.
\ee
Apparently the two answers, (\ref{su2weff1}) and (\ref{su2weff3}),
do not agree for general $X$.  

The resolution of this paradox is that $X$ is {\em not}
a general $4\times4$ complex matrix, but satisfies some
constraints by virtue of its definition (\ref{X}).  In
particular, $X$ can be thought of as the product of two 
antisymmetric matrices ($\Mt BM$ and $\tB$).  Such a matrix,
though not necessarily either symmetric or antisymmetric,
has only half the degrees of freedom that a general matrix,
$Y$, of the same rank would have.

To see this, recall that an appropriate similarity transformation 
$G^{-1}YG$ with $G\in \GL(2n,\C)$ will diagonalize $Y$.  Thus the 
$2n$ eigenvalues of the general rank $2n$ matrix $Y$ are all its 
$\GL(2n,\C)$-invariants.  More properly, a basis of $2n$ independent 
symmetric polynomials of these eigenvalues generates all invariants.  
This basis can conveniently be taken to be $\tr(Y^p)$ for $p=1,\ldots,2n$.
If, on the other hand, $X=AB$ is the product of two antisymmetric 
matrices $A$ and $B$, then a $\GL(2n,\C)$ transformation acts as 
$G^{-1}XG=G^{-1}ABG=G^{-1}AG^{-T}G^TBG$.  We can always choose a 
$G=G_0$ so that $G_0^TBG_0=J$ where $J:=\I_n\otimes i\sigma_2$ is 
the ``unit" skew-diagonal matrix.  
This condition does not fix $G$, since if $H\in\Sp(2n,\C)$ (\ie, 
$H\in\GL(2n,\C)$ satisfies $H^TJH=J$), then $G=G_0H$ will also
give $G^TBG=J$.  For such $G$ we have $G^{-1}XG=H^{-1}A'H^{-T}J$ where
$A'$ is the antisymmetric matrix $A'=G_0^{-1}AG_0^{-T}$.  Now, a
Gram-Schmidt othogonalization argument but with respect to the skew
product defined by $J$ shows that $H$ can always be chosen to
bring $H^{-1}A'H^{-T}$ to skew diagonal form $H^{-1}A'H^{-T}=
{\rm diag}\{a_1,\ldots,a_n\}\otimes i\sigma_2$.  Thus $X$ has 
only $n$ independent (double) eigenvalues,
and a basis of generators of the $\GL(2n,\C)$-invariants of $X$
can be taken to be $\tr(X^p)$ for $p=1,\ldots,n$.  This is half the
number of independent invariants of the general rank $2n$ matrix $Y$,
and implies, in particular, that $\tr(X^p)$ for $p>n$ satisfy additional
identities allowing them to be expressed in terms of products of traces
with $p\le n$.

For example, for $\nf=4$ ($n=2$) the new independent cubic and quartic
identities are easily found to be
\bea\label{rank4idents}
0 &=& 8 \tr X^3 - 6 \tr X^2 \tr X + (\tr X)^3,\nonumber\\
0 &=& 8 \tr X^4 - 4 \tr X^3 \tr X - 2 (\tr X^2)^2  + \tr X^2 (\tr X)^2.
\eea
When $\nf=4$, the identities (\ref{rank4idents}) are easily
checked to imply that $\tr(X^p)=\tr(X'^p)$ for $p=1,\ldots,4$,
where $X':= {1\over2}\tr(X)-X$.  Thus all invariants made from
$X'$ are the same as those for $X$, and in particular, the
two forms of the effective superpotential (\ref{su2weff1}) and
(\ref{su2weff3}) are equivalent.  

\section{$\bf\We$ for $\bf\nf=\nc+2$ superQCD: Seiberg dual analysis}

The IR-equivalent Seiberg dual description \cite{s9411} of
$\SU(\nc)$ superQCD with $\nf=\nc+2$ is $\SU(2)$ superQCD with 
$\nf$ (dual) quarks $q^a_i$ in the fundamental and $\nf$ (dual) 
anti-quarks $\t q^i_a$ in the anti-fundamental, and a set of 
gauge singlets $\h\MM^i_j$ coupled through the superpotential
\be\label{4.24}
\WW=\t q^i_a q^a_j\h\MM^j_i, 
\ee
where $i=1,\ldots,\nf$ and $a=1,2$ are flavor and color indices, 
respectively.  This superpotential breaks the global symmetry of the
dual theory down to $\SU(\nf)\times\SU(\nf)\times\U(1)_B\times\U(1)_R$. 
The dual meson, baryons, and glueball are defined to be
\be
\h\NN^i_j:= \t q^i_a q^a_j,\qquad
\h\BB_{ij} := \e_{ab}q^a_i q^b_j,\qquad
\h{\t\BB^{ij}} := \e^{ab}\t q^i_a \t q^j_b,\qquad
\SS := \tr(w^\a w_\a)/(32\pi^2).
\ee
This dual theory is IR free when $\nf\ge6$ ($\nc\ge4$), and there are 
free quarks, anti-quarks and gluons at the origin of the moduli space.
The $\U(1)_R$ charges are $R(\SS)=2$, $R(\MM)=4/\nf$, and $R(\NN)=R(\BB)
=R(\t\BB)=2\nc/\nf$. 
The chiral ring of the dual theory $(\SS,\MM,\BB,\t\BB)$ is related to 
that of the direct theory $(S,M,B,\tB)$ by \cite{is9509}
\be\label{4.50}
S=-\SS,\qquad 
M=\mu\MM, \qquad
B=i\mu^{-1}\L^{\nf-3}\BB, \qquad
\tB=i\mu^{-1}\L^{\nf-3}\t\BB,
\ee
where $\mu$ is a matching scale defined by
\be
\t\L^{6-\nf} = \mu^\nf \L^{6-2\nf},
\ee
with $\L$ and $\t\L$ being the dynamical scales of the direct and
dual theories, respectively.

The gauge-invariant form of the classical $F$-term equations are
\be\label{dcc1}
\NN = \BB\MM = \MM\t\BB = 0 ,
\ee
where we are using a matrix notation for the fields. 
The $D$-term equations give 
\be\label{dcc2}
\BB\wedge\NN = \NN\wedge\t\BB = \BB\wedge\BB =
\t\BB\wedge\t\BB = \BB\otimes\t\BB - \NN\wedge\NN = 0 ,
\ee
which are the same as the constraints (\ref{2.6}) of $\SU(2)$ 
superQCD discussed in section 2 with the substitution $M\to\NN$.  
The space of solutions to the classical constraints 
(\ref{dcc1}--\ref{dcc2}) has two branches: either $\NN=\BB=\t\BB=0$
and rank$(\MM)>\nc$, or, up to flavor rotations,
\be\label{dcc3}
\NN = 0,\qquad
\MM = \pmatrix{\bf m&&\cr &&\phantom{b}\cr &\phantom{-b}&},\qquad
\BB = \pmatrix{\phantom{\bf m}&&\cr &&b\cr &-b&},\qquad
\t\BB=\pmatrix{\phantom{\bf m}&&\cr &&\tb\cr &-\tb&},
\ee
where $\bf m$ is an $\nc\times\nc$ matrix and $b\tb=0$.

This classical moduli space is clearly not the same as the
moduli space (\ref{4.4}--\ref{cc2}) of the direct $\SU(\nc)$
theory.  However, the dual theory classical constraints 
(\ref{dcc1}--\ref{dcc3}) are expected to be modified quantum 
mechanically by strong coupling effects.  Because even though 
the dual theory is free at the origin of moduli space, arbitrarily 
small vevs for $\MM$ destabilize the free fixed point by giving 
masses to the dual quarks through the superpotential coupling 
(\ref{4.24}), so the theory flows to strong coupling for large 
enough rank of $\MM$.  Strong coupling effects are argued in 
\cite{s9411} to generate the constraints (\ref{4.4}) of the direct 
theory.  

In particular, the branch of (\ref{dcc1}--\ref{dcc3}) 
with rank$(\MM)>\nc$ is lifted, and the $b\tb=0$ constraint
on the other branch is deformed to $b\tb=\det({\bf m})$.
As an example---which will be useful later---of how the
classical constraints are modified by strong coupling effects,
note that if the singlet vev $\MM$ is given a generic value 
(\eg, by constraining it with a Lagrange multiplier term as we 
will do below) with $\BB=\t\BB=0$, the superpotential (\ref{4.24}) 
gives mass to all the dual quarks.  The dual theory thus flows 
to $\SU(2)$ superYang-Mills in the IR with glueball vev $\SS 
= (\t\L^{6-\nf} \det\MM)^{1/2}$ generated by a Veneziano-Yankielowicz 
superpotential 
\be\label{dvy}
\WW_{\rm eff,VY}=2\SS\left(1-\ln\left[
\SS \t\L^{(\nf-6)/2}/\sqrt{\det\MM}\right]\right),
\ee
where $\t\L$ is the strong-coupling scale of the dual theory.
The scale of the glueball appearing this superpotential is 
determined by one loop matching $\t\L_{\rm YM}^6=\t\L^{6-\nf} 
\det\MM$.  Integrating $\SS$ out of (\ref{dvy}) then generates 
an effective superpotential for the singlet field
given by $2 \t\L^{(6-\nf)/2}\sqrt{\det\MM}$, thus lifting the 
rank$(\MM)=\nf$ region of the classical moduli space.

\subsection{Derivation of $\We$}

We now wish to find an exact effective superpotential for the
dual theory which reproduces the strong quantum effects
described in the last two paragraphs.
As in previous sections we start with a tree level superpotential
\be\label{4.25}
\WW_{\rm tree}= \h\NN^i_j\h\MM^j_i + m^i_j(\h\MM^j_i-\MM^j_i)+ 
b^{ij}(\h\BB_{ij}-\BB_{ij})+\tb_{ij}(\h{\t\BB^{ij}}-\t\BB^{ij}),
\ee
where 
\be\label{4.26}
m^i_{\ j}=-{{\del \We} \over  \del {\MM^j_{\ i}}},\qquad
b^{ij}=-{1\over 2}{{\del \We} \over  \del {\BB^{ij}}},\qquad
\tb_{ij}=-{1\over 2}{{\del \We} \over  \del {\tBB}^{ij}},
\ee
are the by-now familiar Lagrange multipliers.  We have only 
specified the vevs of $\h\MM$, $\h\BB$, and $\h{\tBB}$, because 
they completely parametrize the moduli space.

Using the GKA equations (\ref{2.8}) in precisely the same way as 
before, we find a set of equations similar to those, 
(\ref{2.13}--\ref{2.15}), found in section 2,
\be\label{dsu2}
\MM \NN = \SS + 2b\BB,\qquad
\NN \MM = \SS + 2\tBB b,\qquad
\tBB\MMt = -2\NN b,\qquad
\MMt\BB = -2\tb\NN,
\ee
together with the equation coming from the variation of the
singlet field $\MM$:
\be\label{nemm}
\NN=-m.
\ee
Eliminating $\NN$ by plugging (\ref{nemm}) into (\ref{dsu2}) 
gives a set of partial differential equations for the effective 
superpotential
\be\label{4.27}
-\MM m = \SS + 2 b\BB, \qquad
-m\MM = \SS + 2 \tBB\tb, \qquad
\tBB\MMt = 2mb,\qquad
\MMt\BB = 2\tb m.
\ee

To solve for the effective superpotential we multiply the third equation 
in (\ref{4.27}) from the right by $\BB$ and from the left by $\MM$ to 
obtain $\MM\tBB\MMt\BB = 2\MM mb\BB=-2(\SS\I +2b\BB)b\BB$, where in the 
second equality we used the first equation in (\ref{4.27}) to eliminate 
$\MM m$.  Solving this quadratic equation for $b\BB$, we find   
\bea\label{4.29}
b\BB= {1\over4}\left(-\SS+\sqrt{\SS^2-4\MM\tBB\MMt\BB}\right),
&\quad&
\MM m={1\over2}\left(-\SS-\sqrt{\SS^2-4\MM\tBB\MMt\BB}\right),
\nonumber\\
\tBB\tb= {1\over4}\left(-\SS+\sqrt{\SS^2-4\tBB\MMt\BB\MM}\right),
&\quad& 
m\MM={1\over2}\left(-\SS-\sqrt{\SS^2-4\tBB\MMt\BB\MM}\right),
\eea
where the second line comes from solving similar quadratic
equations for $\tBB\tb$ and $m\MM$.

The matrix square roots in the above expressions need some explanation.  
In order to make sense of them, consider a region in the configuration 
space where the magnitude of each element in the matrix $\MM\tBB\MMt\BB$ 
is much smaller than $\SS^2$.  We can then expand the the square root 
as a power series, $\sqrt{\SS^2-4\MM\tBB\MMt\BB} = \SS\sqrt{\I} (\I-
2\SS^{-2}\MM\tBB\MMt\BB+\cdots)$, and by analytic continuation we extend 
the result to include all points on the parameter space.  However, for 
this to be a definition of the square root, we still need to determine 
$\sqrt{\I}$.  In general $\sqrt{\I}=2P-\I$ where $P$ can be any projection 
matrix ($P^2=P$).  Which $P$ should we use?  The following argument shows 
that we have to take $\sqrt{\I}=\pm\I$.  As we saw in the discussion 
surrounding (\ref{dvy}), we expect to generate a non-zero $\SS$ at points 
in the parameter space where $\MM\neq0$ but $\BB=\tBB=0$, so this is a 
suitable region to evaluate the square roots.  At such  points (\ref{4.27})
implies that $m\MM=\MM m = -\SS$ and $\tBB\MMt=\MMt\BB=0$.  But this is
only consistent with (\ref{4.29}) if $\sqrt{\SS^2\I}= \SS\I$ implying that 
$\sqrt{\I}=\pm\I$.

It is straightforward to integrate (\ref{4.29}) for the effective 
superpotential to get
\bea\label{4.33}
\We &=& {\SS\over4} \ln\det\left(\sqrt{\SS^2-4\MM\tBB\MMt\BB}
-\SS \over \sqrt{\SS^2-4\MM\tBB\MMt\BB}+\SS \right)
-{\SS\over4}\ln\det(\t\L^{-6}\MM\tBB\MMt\BB)
+ \SS\ln\det(\t\L^{-1}\MM)  \nonumber\\
&& \mbox{} + {1\over2} \tr\sqrt{\SS^2-4\MM\tBB\MMt\BB}
+ {1\over2}(4-\nf)\SS\left[\a-\ln(\SS/\t\L^3)\right]. 
\eea
We used the $\U(1)_R$ symmetry to fix the $\SS$ dependence up to an
undetermined integration constant $\a$.  Evaluating $\We$ at $\BB=
\tBB=0$ gives, after a somewhat delicate cancellation,
\be\label{webz}
\We(\BB=\tBB=0) = 2\SS\left(\a+(1-\a){\nf\over4}
-\ln\left[\SS\t\L^{(\nf/2)-3}/\sqrt{\det\MM}\right]\right) .
\ee
Comparing to the answer (\ref{dvy}) expected from 
the strong coupling analysis, fixes $\a=1$.
Note that this limiting form (\ref{webz}) is already a
check that the effective superpotential (\ref{4.33}) is
consistent with the quantum modified constraints.  In particular
the $\SS\ln\det\MM$ term in (\ref{webz}) serves to lift the 
whole rank$\MM=\nf$ region of the classical moduli space
(\ref{dcc1}--\ref{dcc3}), in accordance with the expected 
quantum constraints (\ref{4.4}--\ref{cc2}).

So, our final result for the effective superpotential
for the Seiberg dual theory is
\bea\label{dweff}
\We &=& {\SS\over4} \ln\det\left(\sqrt{\SS^2-4\MM\tBB\MMt\BB}
 -\SS \over \sqrt{\SS^2-4\MM\tBB\MMt\BB}+\SS \right)
 -{\SS\over4}\ln\det\left(\MM\tBB\MMt\BB\right)
 +\SS\ln\det\MM \nonumber\\
&& \mbox{} + {1\over2} \tr\sqrt{\SS^2-4\MM\tBB\MMt\BB}
 + (4-\nf){\SS\over2}(1-\ln\SS) + (6-\nf)\SS\ln\t\L .
\eea
Using the chiral ring mappings (\ref{4.50}), this can be
expressed in terms of the fields of the direct theory.  Note
that since $S=-\SS$, the definition of the branch of
the square root made above, $\sqrt{\SS^2\I}=\SS\I$, now
becomes $\sqrt{S^2\I}=-S\I$.  We make this minus sign
explicit by changing the signs of all square roots and
keeping the convention $\sqrt{S^2\I}=+S\I$.  We then find
\be\label{weffws}
\we = {s\over4} \ln\left[\det\left( \sqrt{s^2+4x}+s \over 
\sqrt{s^2+4x}-s \right) {\det x\over \det^4 M}\right]
-{1\over2}\tr\sqrt{s^2+4x} +(\nf-4){s\over2}\left[1-\ln s\right],
\ee
where we have defined the shorthands
\be\label{shorthand}
\we := \L^{\nf-3}\We, \qquad s:=\L^{\nf-3} S ,\qquad x:=M\tB\Mt B ,
\ee
to avoid having to write factors of $\L$.  This is the effective 
superpotential for $\SU(\nc)$ superQCD with $\nf=\nc+2$.

\subsection{Integrating out the glueball field}

Away from the origin of moduli space, the glueball $s$ is expected 
to be massive.  Solving its equation of motion, $\del\we/\del s=0$, 
gives $s=s_*(M,B,\tB)$ where $s_*$ is defined implicitly by
\be\label{sstar}
s_*^{2(\nf-4)} = \det\left(\sqrt{s_*^2+4x}+s_*\over 
\sqrt{s_*^2+4x}-s_*\right) {\det x\over \det^4 M } .
\ee
Substituting this in $\we$ gives the effective superpotential as 
a function of meson and baryons only,
\be\label{weffss}
\we|_{s_*}
= -2s_* \left[1+{1\over4}\tr\left(\sqrt{1+4{x\over s_*^2}}
-1\right)\right] .
\ee
The relation (\ref{sstar}) gives $s_*$ as a complicated
function of $M$, $B$, and $\tB$, which makes it difficult
to deduce the equations of motion for these fields from $\we$.  

Before solving (\ref{sstar}) for $s_*$, we can extract some of
the constraints that $M$, $B$, and $\tB$ must satisfy on the
moduli space.  Since $\we$ is singular, as shown in \cite{ae0510}
it must be regularized in order to extract its physical predictions.
The idea behind the regularization is to deform $\We$ by introducing 
some regularizing parameters $\mu$, $\b$, $\tbe$ as follows
\be\label{4.40}
\we \to \we^{\mu,\b,\tbe} = \we + \mu^i_j M^j_i 
+ \b^{ij} B_{ij} +\tbe_{ij}\tB^{ij} .
\ee
Generic small values of $\mu$, $\b$, and $\tbe$, will fix $M$, $B$, 
and $\tB$ to some values which will be shifted from the moduli 
space (the extrema of $\we$) by positive powers of the regularizing
parameters.  Thus as $\mu,\b,\tbe\to0$, $\{M,B,\tB\}$ will approach
some point on the moduli space.
However, the specific point reached on the moduli space will
depend on how the regularizing parameters scale to zero:  as
different orders of limits of $\mu,\b,\tbe\to0$ are taken, the
whole moduli space will be scanned.  (Note that some vanishing
limits of the regularizers can also send $M$, $B$, or $\tB$ to 
infinity; this is unavoidable since the moduli space itself
stretches off to infinity.)

This regularization procedure should be compared with the trick we 
used of introducing Lagrange multipliers (\ref{4.25}) to 
derive equations for $\we$ from the GKA equations.  With the
replacement of the Lagrange multipliers with the regularizing
parameters, $\{m,b,\tb\}\to\{\mu,\b,\tbe\}$, the identification
of the Lagrange multipliers as derivatives of $\we$, (\ref{4.26}), 
is now interpreted as the regularized equations of motion.  Thus, 
extremizing $\we^{\mu,\b,\tbe}$ using (\ref{weffws}) gives
\bea\label{4.41}
4\b B  = s-\sqrt{s^2+4x}, &\qquad&
2M\mu  = s+\sqrt{s^2+4x}, \nonumber\\
4\tB\tbe = s-\sqrt{s^2+4y}, &\qquad&
2\mu M = s+\sqrt{s^2+4y}, 
\eea
where $y:=M^{-1}xM$.  (\ref{4.41}) is equivalent to (\ref{4.29}), 
the original differential 
equations---with the operator mappings (\ref{4.50})---we integrated 
to get $\we$ in the first place.  Then, reversing the manipulations 
which led from (\ref{4.27}) to (\ref{4.29}) gives
\be\label{cc1}
M\mu=s-2\b B, \quad \mu M=s-2\tB\tbe,\quad
\tB\Mt = -2\mu\b, \quad \Mt B = -2\tbe\mu.
\ee
These show that independent of how $\mu,\b,\tbe\to0$, we always end 
up with $\tB\Mt=\Mt B=0$.  This implies in particular that
$x = -4 \b\mu^T\tbe\mu$, and so
also always vanishes with the regularizing parameters.  Also, if as 
$\mu,\b,\tbe\to0$ $M$, $B$, and $\tB$ remain finite, then $s\to0$ 
as well.  

This shows that the extrema of the superpotential (\ref{weffws}) 
satisfy the constraints $\tB\Mt=\Mt B=s=0$, which are, indeed, part of 
the constraints (\ref{4.4}) describing the $\nf=\nc+2$ superQCD moduli 
space.  The remaining constraints should also follow from the effective
superpotential.  However, they are much harder to derive, as they 
require solving for $s_*$ in (\ref{sstar}).  Now the argument of the
last paragraph shows that $s_*=0$, but, because of the singular
nature of the superpotential, it is incorrect to simply plug this
value into (\ref{weffss}) to find $\we$.  Instead, $s_*$ should be found 
for generic $M$, $B$, and $\tB$, and then the extrema of the resulting
effective superpotential can be analyzed by regularizing it, as above.

The equation (\ref{sstar}) can be solved systematically for $s_*$
by assuming that (all the eigenvalues of) $\xi := x/s_*^2 \ll 1$ and
expanding in powers of this parameter.  The leading order solution
is $s_*^2=\det M$, which can be checked to be consistent with the
assumption that $||\xi|| \ll 1$.  Change variables to 
\be
\s^2 := {s^2_* \over \det M},
\qquad
X := {x\over\det M} ,
\ee
so that (\ref{sstar}) and (\ref{weffss}) become
\bea\label{sig}
\s^{2(\nf-4)} &=& \det X\,\det\left(\sqrt{\s^2+4X}+\s\over 
\sqrt{\s^2+4X}-\s\right) 
=  2^{-2\nf}\,\det\left(\sqrt{\s^2+4X}+\s\right)^2 ,
\\ \label{weff}
\we &=& -2\sqrt{\det M}\left[\s+{1\over4}\tr\left(\sqrt{\s^2+4X}
-\s\right)\right] .
\eea
Expand the right side of (\ref{sig}) in powers of $X/\s$ and 
solve it consistently order-by-order in a power series
expansion $\s^2 = 1+ \cdots$ to get
\bea\label{sexp}
\s^2 &=& 1 - {\tr X\over2} - {(\tr X)^2\over8} + {3\tr(X^2)\over4}  
- {(\tr X)^3\over12} + {3\tr X \tr(X^2)\over4} - {5\tr(X^3) \over3} 
\\ && \ \ \mbox{}
- {9(\tr X)^4\over 128} + {27(\tr X)^2\tr(X^2)\over32}
- {27\tr(X^2)^2\over32} - {5\tr X\tr(X^3)\over2}
+ {35\tr(X^4)\over8}
+ {\cal O}(X^5).\nonumber
\eea
Plugging this into (\ref{weff}) gives
\bea\label{weffexp}
\we &=& -2\sqrt{\det M} \Bigl[ 
1 + {\tr X\over4} + {(\tr X)^2\over32} - {\tr(X^2)\over8}  
+ {5(\tr X)^3\over384} - {3\tr X \tr(X^2)\over32} + {\tr(X^3)\over6}
\nonumber\\ && \qquad\qquad\quad \mbox{}
+ {49(\tr X)^4\over6144}  - {21(\tr X)^2 \tr(X^2)\over256} 
+ {9\tr(X^2)^2\over128} + {5\tr X\tr(X^3)\over24}
- {5\tr(X^4)\over16} 
\nonumber\\ && \qquad\qquad\quad \mbox{}
+ {\cal O}(X^5) \Bigr].
\eea
This is the answer, to order $X^4$, for the effective superpotential
for $\nf=\nc+2$ superQCD found by integrating the Seiberg dual
GKA equations and integrating out the glueball.

Alternatively, we can derive a differential equation satisfied
by $\we$ as a result of integrating out $s$.
Two of the equations of motion that we originally integrated
to get the superpotential (\ref{4.41}) become, when written in 
terms of $\s$ and $X$,
\bea\label{eom1}
4\b B  &=& \sqrt{\det M} \left(\s-\sqrt{\s^2+4X}\right),
\nonumber\\
2M\mu  &=& \sqrt{\det M}\left(\s+\sqrt{\s^2+4X}\right).
\eea
On the other hand, upon integrating out $s$, we have seen that 
$\we$ takes the form
\be\label{weff2}
\we = \sqrt{\det M} \, f(X)
\ee
for some function $f$.  (This form just follows from the symmetries.)
Two of the equations of motion following from this form of $\we$ are
\bea\label{eom2}
4\b B  &=& \sqrt{\det M} \left(\phantom{+}4Xf'\ \right),
\nonumber\\ 
2M\mu  &=& \sqrt{\det M}\left(-4Xf'-f+2\tr(Xf')\right),
\eea
where $f'$ is the matrix derivative $df/dX$.

Equating and adding (\ref{eom1}) and (\ref{eom2}) gives
\be\label{sigeq}
\s = -{1\over2}f+\tr(Xf') ,
\ee
while equating and multiplying (\ref{eom1}) and (\ref{eom2}) gives
\be\label{mde}
1=f'\left(f-2\tr(Xf')+4Xf'\right).
\ee
This is a nonlinear first order (matrix) differential equation
for $\we$ with the glueball integrated out.  We do not 
know how to integrate this equation in closed form when rank$(X)>1$.  
But it is straightforward to check that a series expansion of
the solution to (\ref{mde}) with boundary condition $f(0)=-2$
reproduces (\ref{weffexp}).

It may be clarifying to note that (\ref{mde}) has a one-parameter 
family of solutions.  For if $f(X)$ is a solution, then so is 
\be\label{deqsymm}
f_{(a)}(X) := af(a^{-2} X)
\ee
for any $a\in\C^*$.  By (\ref{sigeq}) this implies that
$\s(X)$ changes to $a\s(a^{-2} X)$;  but it is easy to
check that the $\s$ equation of motion (\ref{sig}) does
{\em not} have this symmetry.  The algebraic equation (\ref{sig}) 
determining $\s$ has more information than the differential
equation (\ref{mde}), and so picks out a single instance of
the family of solutions (\ref{deqsymm}), namely the one obeying
the boundary condition $f(0)=-2$.  (For example, 
$-2\tr(\sqrt{2X})$ solves (\ref{mde}) but does not satisfy 
(\ref{sig}), so is not a physical solution.)

\subsection{Comparing to the direct result when $\bf\nf=4$}

Is (\ref{weffexp}) the correct superpotential?  A basic check
is to see whether its extrema (computed by appropriately
regularizing, as explained above) reproduce the moduli space
given by the constraints (\ref{4.4}).  This seems a very difficult
check to perform since we do not have a closed analytic form for $\we$.

However, there is one case where we can carry out a non-trivial
check.  When $\nf=4$, $\nc=2$, so in this case we should reproduce 
the superpotential (\ref{su2weff1}) found in sections 2 and 3 for 
the $\SU(2)$ theory.  Expanding (\ref{su2weff1}) in powers of
$X$, we find
\bea\label{SU2weffexp}
\L{\cal W}_{\rm eff, SU(2)} &=& -2 \sqrt{\det M} \Bigl[ 
1 + {\tr X\over4} + {(\tr X)^2\over32} - {\tr(X^2)\over8}  
+ {(\tr X)^3\over384} - {\tr X \tr(X^2)\over32} + {\tr(X^3)\over12} 
\nonumber\\ && \qquad\qquad\quad \mbox{}
+ {(\tr X)^4\over6144} - {(\tr X)^2 \tr(X^2)\over256}  
+ {\tr(X^2)^2\over128}  + {\tr X \tr(X^3)\over48} 
- {\tr(X^4)\over16} 
\nonumber\\ && \qquad\qquad\quad \mbox{}
+ {\cal O}(X^5) \Bigr].
\eea
Though it does not coincide with the expansion (\ref{weffexp}) starting 
at order $X^3$, we must bear in mind the identities (\ref{rank4idents})
that traces of powers of $X$ satisfy starting at cubic order.  One 
finds that the difference between (\ref{SU2weffexp}) and 
(\ref{weffexp}) is proportional to these identities, and so 
vanishes.  Thus the effective superpotential found by integrating 
the dual GKA equations matches the correct result at least to 
quartic order in $X$.  

\section*{Acknowledgments}
It is a pleasure to thank C. Beasely, M. Douglas, S. Hellerman, 
N. Seiberg, P. Svr\v cek and E. Witten for helpful comments and 
discussions, and to thank the School of Natural Sciences at the 
Institute for Advanced Study for its hospitality and support.  
PCA and ME are supported in part by DOE grant DOE-FG02-84ER-40153.  
PCA was also supported by an IBM Einstein Endowed Fellowship.


\begin{thebibliography}{99}

\bibitem{is9509}
K.A. Intriligator and N. Seiberg, 
{\sl Lectures on supersymmetric gauge theories and electric-magnetic 
duality}, 
\npps{45BC}{1996}{1}, [\hepth{9509066}].

\bibitem{p9702}
M. E. Peskin, 
{\sl Duality in supersymmetric Yang-Mills theory},
[\hepth{9702094}].

\bibitem{ads84}
I.~Affleck, M.~Dine and N.~Seiberg, 
{\sl Dynamical supersymmetry breaking in supersymmetric QCD},
\npb{241}{1984}{493}; 
{\sl Dynamical supersymmetry breaking in four dimensions and its
phenomenological implications},
\npb{256}{1985}{557}.

\bibitem{nsvz85}
V. A. Novikov, M. A. Shifman, A. I. Vainshtein and V. I.
Zakharov, {\sl Supersymmetric instanton calculus: gauge theories 
with matter}, 
\npb{260}{1985}{157}.

\bibitem{s9402}
N. Seiberg, 
{\sl Exact results on the space of vacua of four dimensional 
SUSY gauge theories},
\prd{49}{1994}{6857}, [\hepth{9402044}].

\bibitem{s9411}
N. Seiberg, 
{\sl Electric-magnetic duality in supersymmetric nonabelian 
gauge theories}, 
\npb{435}{1995}{129}, [\hepth{9411149}].

\bibitem{ae0510}
P. C. Argyres and M. Edalati, 
{\sl On singular effective superpotentials in supersymmetric 
gauge theories}, 
[\hepth{0510020}].

\bibitem{bw0409}
C. Beasley and E. Witten,
{\sl New instanton effects in supersymmetric QCD},
\jhep{0501}{2005}{056}, [\hepth{0409149}].

\bibitem{c0305}
S. Corley,
{\sl Notes on anomalies, baryons, and Seiberg duality},
[\hepth{0305096}]

\bibitem{cdsw0211}
F. Cachazo, M. R. Douglas, N. Seiberg and E. Witten,
{\sl Chiral rings and anomalies in supersymmetric gauge theory}, 
\jhep{0212}{2002}{071}, [\hepth{0211170}].

\bibitem{s0212}
N. Seiberg,
{\sl Adding fundamental matter to `Chiral rings and anomalies in 
supersymmetric gauge theory'},
\jhep{0301}{2003}{061}, [\hepth{0212225}].

\bibitem{binor0303}
A. Brandhuber, H. Ita, H. Nieder, Y. Oz and C. Romelsberger, 
{\sl Chiral rings, superpotentials, and the vacuum structure of 
$\NN=1$ supersymmetric gauge theories}, 
\atmp{7}{2003}{269}, [\hepth{0303001}].

\bibitem{s0311}
P. Svr\v cek,
{\sl On non-perturbative exactness of Konishi anomaly and the 
Dijkgraaf-Vafa conjecture},
\jhep{0410}{2004}{028}, [\hepth{0311238}].

\bibitem{ils9403}
K. A.~Intriligator, R. G.~Leigh and N.~Seiberg,
{\sl Exact superpotentials in four-dimensions},
\prd{50}{1994}{1092}, [\hepth{9403198}].

\bibitem{i9407}
K.~A.~Intriligator,
{\sl 'Integrating in' and exact superpotentials in 4-d},
\plb{336}{1994}{409}, [\hepth{9407106}].

\bibitem{vy82}
G. Veneziano and S. Yankielowicz,
``An effective lagrangian for the pure N=1 supersymmetric 
Yang-Mills theory,'' 
\plb{113}{1982}{231}.

\end{thebibliography}
\end{document}